\documentclass[12pt]{article}
\usepackage{graphicx}
\begin{document}
\rightline{NKU-2013-SF5}
\bigskip

\newcommand{\be}{\begin{equation}}
\newcommand{\ee}{\end{equation}}
\newcommand{\noi}{\noindent}
\newcommand{\refb}[1]{(\ref{#1})}
\newcommand{\ra}{\rightarrow}

\begin{center}
{\Large\bf Nariai black holes with quintessence}

\end{center}
\hspace{0.4cm}
\begin{center}
Sharmanthie Fernando \footnote{fernando@nku.edu}\\
{\small\it Department of Physics \& Geology}\\
{\small\it Northern Kentucky University}\\
{\small\it Highland Heights}\\
{\small\it Kentucky 41099}\\
{\small\it U.S.A.}\\

\end{center}

\begin{center}
{\bf Abstract}
\end{center}

In this paper we study the properties of  Schwarzschild black hole surrounded by quintessence matter. The main objective of the paper is to show the existence of Nariai type black hole for special values of the parameters in the theory.  The Nariai black hole with the quintessence has the topology $dS_2 \times S_2$ with $dS_2$ with a different scalar curvature than what  would be expected for the Schwarzschild-de Sitter degenerate black hole. Temperature and the entropy for the Schwarzschild-de Sitter black hole and the Schwarzschild-quintessence black hole  are compared. The temperature and the curvature are computed for general values of the state parameter $\omega$.

\hspace{0.7cm}

{\it Key words}: static, quintessence, Nariai, black hole

\section{ Introduction}

There is overwhelming observational evidence that our universe is undergoing accelerated expansion \cite{perl} \cite{riess} \cite{sper}  \cite{teg} \cite{sel}. Such a behavior is consistent with a dark energy density which is constant. An interesting review on dark energy is given in \cite{sami}.
One of the simplest candidates for the dark energy density is the cosmological constant with a state parameter $\omega = -1$. However,  there are major problem that is yet to be understood about the cosmological constant from a fundamental physics point of view. The observed value is too small in comparison with the theoretical prediction which is known as the fine-tuning problem \cite{wein}.

There are alternative models  as candidates for dark energy. Most of these models consist of  dynamical scalar fields. Such models  includes but not limited to, quintessence \cite{carroll}, chameleon fields \cite{khoury}, K-essence \cite{amer},   tachyon field \cite{pad}, phantom dark energy \cite{cald} and dilaton dark energy \cite{gas}.  A detailed description of various scalar field models can be found in the review by Copeland \cite{copeland}.

In this paper we will focus on black holes surrounded by quintessence matter. The quintessence is a scalar field coupled to gravity and its  state  parameter $\omega$ larger than -1. A nice review about the quintessence is in \cite{shin}. There are many works related to the quintessence in the literature. We will only provide a handful of works here.
The correspondence between the quintessence and the tachyon dark energy models with a constant equation of state was studied by Avelino et.al \cite{ave}. Singularity of spherically symmetric space-time in quintessence/phantom dark energy universe was studied by Nojiri and Odintsov \cite{noj}. A discussion on whether we will ever distinguish between quintessence and the cosmological constant was given in \cite{chon}. Dynamics of interacting phantom and quintessence dark energy is given in \cite{umar}.

The flow of the paper is as follows. In section 2, an introduction to black holes surrounded by the quintessence is given. In section 3, black holes surrounded by the quintessence with a state parameter  $\omega = -\frac{2}{3}$ is presented in detail. In section 4, we give some details about the Schwarzschild-de Sitter black hole.  A comparison of the black holes surrounded by the quintessence and the one with the cosmological constant
 is done in section 5. In section 6, the Nariai black hole with quintessence is studied for $\omega = -\frac{2}{3}$ and extended to general $\omega$ values in section 7.  Finally, the conclusion  is given in section 8.

\section { Black holes surrounded by the quintessence}

In this section we  will  introduce the Schwarzschild black hole surrounded by the quintessence derived  by Kiselev \cite{kiselev}.  His derivation assumed a static spherically symmetric gravitational field with the energy momentum tensor,

\be
T^t_t = T^r_r = \rho_q
\ee
\be
T^{\theta}_{\theta} = T^{\phi}_{\phi} = -\frac{ 1}{2} \rho_q  ( 3 \omega + 1)
\ee
Here $\rho_q$ is the density of quintessence matter given by,
\begin{equation} \label{rho}
\rho_q = - \frac{ \alpha}{ 2} \frac { 3 \omega}{ r^{ 3 ( 1 + \omega)}}
\end{equation}
and $\omega$ is the quintessential state parameter.

By solving the Einstein equations  with the above energy-momentum tensor, a black hole solution with the following metric was obtained,
\begin{equation}
ds^2 = - f(r) dt^2 + \frac{ dr^2}{ f(r)} + r^2 ( d \theta^2 + sin^2 \theta d \phi^2)
\end{equation}
where,
\begin{equation}
f(r) = 1 - \frac{ 2 M} { r} - \frac{ \alpha}{ r^{ 3 \omega + 1}}
\end{equation}
Here $M$ is the mass and,  $ \alpha$ a normalization factor. In order to obtain a cosmological horizon similar to the Schwarzschild-de Sitter black hole, the parameter $\omega$ has to have the  range,
\begin{equation}
-1 < \omega < - \frac{ 1}{ 3}
\end{equation}
In this paper, we will only consider $\omega$ for these ranges of values.

The quintessence matter has the equation of state as,
\begin{equation}
p_q = \omega  \rho_q
\end{equation}
Here $p_q$ which is  the pressure  has to be negative to cause acceleration. If we require   the matter energy density $\rho_q$  to be positive,  the parameter $ \alpha$ has to be positive as evident  from eq$\refb{rho}$. The basis  for the choices of the parameters and for  details on the derivation of the metric, reader is referred to the original paper of Kiselev \cite{kiselev}. In Fig.1, the function $f(r)$ is plotted for various values of $\omega$. Smaller $\omega$ leads to a smaller cosmological horizon for fixed mass and $\alpha$.

\begin{center}
\scalebox{.9}{\includegraphics{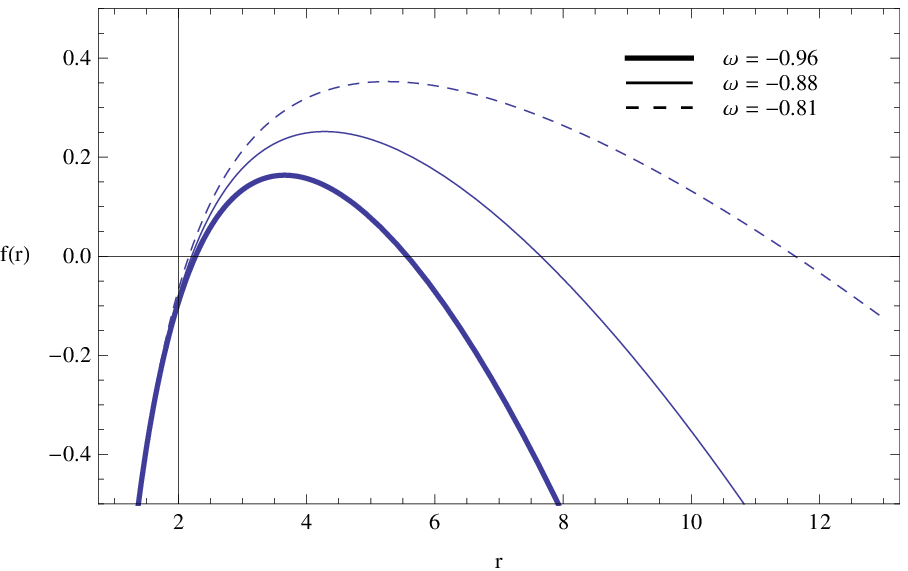}}

\vspace{0.3cm}

 \end{center}

Figure 1. The figure shows the $f(r)$ vs $r$ with three values of $\omega$. Here, $ M = 1$ and $ \alpha = 0.025$.\\

\noindent
When $ \omega = -1$, the function $f(r)$ reduces to,
\begin{equation}
f(r) = 1 - \frac{ 2 M} { r}   - \alpha r^2
\end{equation}
which is the Schwarzschild-de Sitter black hole (one can replace $\alpha = \frac{ \Lambda}{3}$ where $ \Lambda$ is the cosmological constant).

The scalar  curvature of the metric is given by,
\be
R = \frac {3 \alpha \omega ( 1 - 3 \omega)}{ r^{ 3( \omega + 1)}}
\ee
There is a singularity at $ r =0$ except for  $ \omega = \{ -1, 0, -\frac{1}{3} \}$.

\section{ Schwarzschild black hole surrounded by the quintessence for $\omega = -\frac{2}{3}$}

In this paper our main focus will be on the black hole surrounded by quintessence with the parameter $ \omega =- \frac{2}{3}$. For such a value, the metric takes form,
\begin{equation}
ds^2 = -  ( 1 - \frac{ 2 M}{r} -\alpha r ) dt^2 + \frac{ dr^2}{ ( 1 - \frac{ 2 M}{r} - \alpha r)} + r^2 ( d \theta^2 + sin^2 \theta d \phi^2)
\end{equation}
The black hole has two horizons at,
\begin{equation}
r_{b} = \frac { 1 - \sqrt{ 1 - 8 \alpha M} }{ 2 \alpha}
\end{equation}
and,
\begin{equation}
r_{c } = \frac { 1 + \sqrt{ 1 - 8 \alpha M} }{ 2 \alpha}
\end{equation}

\begin{center}
\scalebox{.9}{\includegraphics{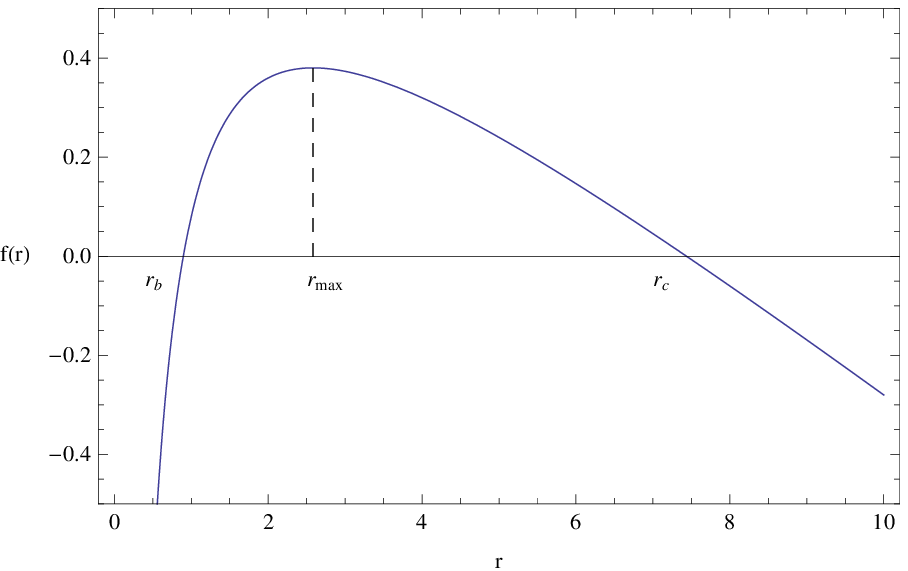}}

\vspace{0.3cm}

 \end{center}

Figure 2. The figure shows the $f(r)$ vs $r$ with $r_{max}$. Here, $ M = 0.4$ and $ \alpha = 0.12$.\\

\noindent
Here $r_b$ is the black hole event horizon similar to the one in the Schwarzschild black hole and $r_{c}$ is the cosmological horizon. For the Schwarzschild black hole with the quintessence matter, the horizons are possible only if 
\begin{equation}
1 - 8 \alpha M > 0 \Rightarrow  M < \frac{ 1}{ 8\alpha}
\end{equation}
When $ M = \frac{ 1}{ 8 \alpha}$, one obtain degenerate horizons at,
\begin{equation}
r_b = r_{c} = \frac{ 1}{ 2 \alpha}
\end{equation}
Hence, for fixed $\alpha$, there is a maximum mass 
$M_{max} = \frac{ 1}{ 8 \alpha}$ for which horizons exists. This is shown in Fig.3. When $ 0 < M < M_{max}$, there is a 
static region between $r_b$ and $r_{c}$ as shown in Fig.2.

\begin{center}
\scalebox{.9}{\includegraphics{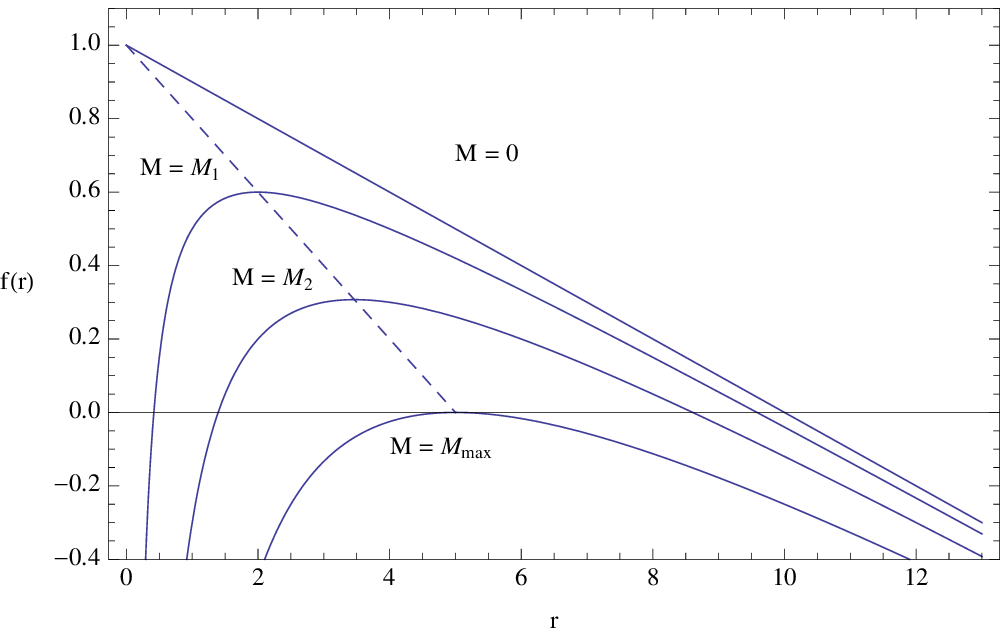}}

\vspace{0.3cm}

 \end{center}

Figure 3. The function $f(r)$ is plotted for various values of $M$ with fixed $\alpha$. The doted curve shows the maximum point of $f(r)$ for various values of $M$. Here $ M_1 < M_2 < M_{max}$. \\

\noindent
For fixed mass $M$, there is a maximum value of $\alpha$ where the horizons are possible. This is shown in the Fig.4.

\begin{center}
\scalebox{.9}{\includegraphics{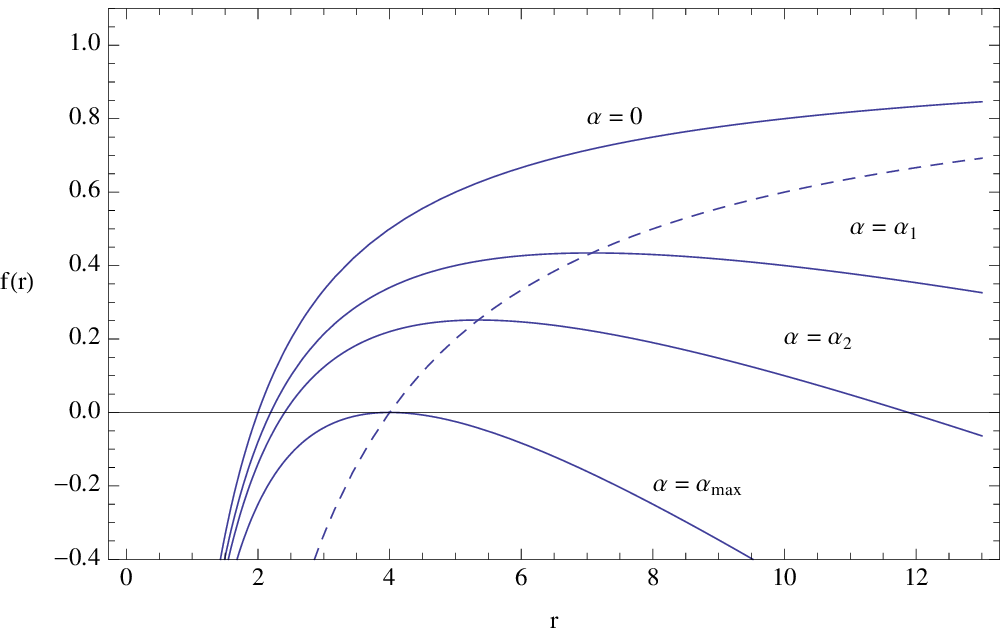}}

\vspace{0.3cm}

 \end{center}

Figure 4. The figure shows the $f(r)$ vs $r$ for various values of $\alpha$ and for fixed $M$.\\

\noindent
The magnitude of the acceleration of a particle in the background of the Schwarzschild black hole with the quintessence is given by,
\begin{equation}
a = \sqrt{ a^{\mu} a_{\mu}} = 
\frac{\left| \frac{ df}{ dr} \right|}{ \sqrt{ 2 f'(r)}}
\end{equation}
Hence, when $f(r)$ has the maximum, the acceleration, $a =0$. i.e., $f'(r) =0$. The corresponding location is given by $ r_{max} = \sqrt{ \frac{ 2 M}{ \alpha}}$ as represented in the Fig.2. Hence, $f(r_{max}) =1 - \sqrt{ 8 M \alpha}$.\\

Since the black hole inconsideration is similar to asymptotically de Sitter space-time containing a cosmological horizon, there are two temperatures defined locally at each horizon. Hence,  the Hawking temperature of the Schwarzschild-quintessence black hole at each horizon is given by,
\begin{equation}
T_{b} = \frac{ 1}{ 4 \pi} \left| \frac{ 2 M}{ r_b^2} - \alpha \right|
\end{equation}

\begin{equation}
T_{c} = \frac{ 1}{ 4 \pi} \left| \frac{ 2 M}{ r_c^2} - \alpha \right|
\end{equation}

There are several works that have focused on Schwarzschild black hole surrounded by the quintessence. Thermodynamics of this black hole is studied in \cite{thara}. Null geodesics and gravitational lensing has been studied in \cite{fernando1}. Late time tails of fields around such a black hole was studied by Varghese and Kuriakose in \cite{varg}. The influence of the quintessence on light deflection was studied in \cite{liu}.  In \cite{chen},  black hole solutions surrounded by quintessence  for $d$ dimensions with a negative cosmological constant were  presented  and  studies were done on holographic superconductors. Hawking radiation by $d$-dimensional static spherically symmetric black holes surrounded by quintessence were studied by Chen et.al \cite{chen2}.

\section{ Schwarzschild-de Sitter black hole}

Let us write the Schwarzschild-de Sitter black hole metric here, for the purpose of comparison.
\begin{equation}
ds^2 = -  ( 1 - \frac{ 2 M}{r} -\frac{\Lambda}{3} r^2 ) dt^2 + \frac{ dr^2}{ ( 1 - \frac{ 2 M}{r} - \frac{ \Lambda}{3} r^2)} + r^2 ( d \theta^2 + sin^2 \theta d \phi^2)
\end{equation}
For $ 0 < 9 \Lambda M^2 < 1$, there are two horizons, $r_b$ (the black hole event horizon) and $r_{c}$ (the cosmological horizon). They are given by,
\begin{equation}
r_b = \frac{ 2}{ \sqrt{\Lambda}} cos \left( \frac{\gamma}{3} + \frac{ 4 \pi}{3} \right)
\end{equation}
\begin{equation}
r_{c} = \frac{ 2}{ \sqrt{\Lambda}} cos \left( \frac{\gamma}{3} \right)
\end{equation}
where,
\begin{equation}
\gamma = cos^{-1}\left( - 3 M \sqrt{\Lambda} \right)
\end{equation}
The global structure of the Schwarzschild-de Sitter black hole is discussed in \cite{raphael}. 
The Hawking temperature of the Schwarzschild-de Sitter black hole at the two horizon are given by,
\begin{equation}
T_{b} = \frac{ 1}{ 4 \pi} \left| \frac{ 2 M}{ r_b^2} - \frac{2}{3}\Lambda r_b \right|
\end{equation}
\begin{equation}
T_{c} = \frac{ 1}{ 4 \pi} \left| \frac{ 2 M}{ r_c^2} -\frac{2}{3} \Lambda r_c \right|
\end{equation}
When $ 9 \Lambda M^2 =1$, the space-time has degenerate horizons at $ r = 3 M= \frac{1}{\sqrt{\Lambda}}$. This is the so called Nariai black hole \cite{perry}\cite{cho}. The temperature between the black hole horizon and the cosmological horizon are the same and  is given by $T = \frac{ \sqrt{\Lambda}}{2 \pi}$. The evaporation of such black holes were investigated by Bousso and Hawking in \cite{raphael2}. The proper distance between the horizons are not zero. It is given by 
$\frac{ \pi}{ \sqrt{\Lambda}}$ \cite{eune}. There are several works that have discussed the thermodynamical properties of the Nariai black  hole \cite{eune}\cite{eune2}\cite{yun}. There are three  papers we like to mention here with regard to Nariai black holes in general. Generalized Nariai solutions for Yang-type monopoles were studied by Diaz and Segui \cite{diaz}.  Charged Nariai black holes with a dilaton were studied by Bousso \cite{raphael3}. Cold, ultracold and Nariai black holes for charged  black holes with the quintessence were studied by the current author in \cite{fernando3}.

In the degenerate horizon  case, the metric function f(r) becomes
\begin{equation}
f(r)_{Sch-de~Sitter} = \frac{ -1 }{ 27 M^2 r} ( r - 3 M)^2 ( r + 6 M)
\end{equation}
In this case,  $f(r)<0$ for all positive $r$. Therefore $r$ becomes the time coordinate and $t$ becomes a spatial coordinate. The structure of the extreme Schwarzschild black hole was studied in \cite{podolsky}.  More details on Schwarzschild-de Sitter black hole is given in the  book by Griffiths and Podolsky \cite{pod}

The metric function for the Schwarzschild-de Sitter black hole is plotted  in Fig.5.

\begin{center}
\scalebox{.9}{\includegraphics{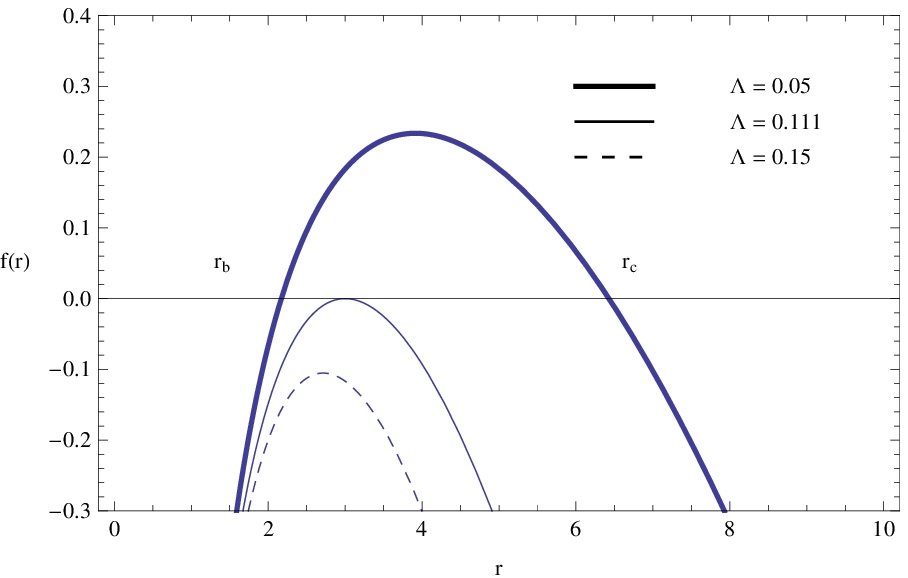}}

\vspace{0.3cm}

 \end{center}

Figure 5. The figure shows  $f(r)$ vs $r$ for various values of $\Lambda$ and for fixed $M$.\\


\section{ Comparison of the Schwarzschild black hole with the quintessence and the Schwarzschild-de Sitter black hole}

Now, the two metric functions for Schwarzschild-quintessence and Schwarzschild-de Sitter black holes are plotted in Fig.6 for the sake of comparison. 

\begin{center}
\scalebox{.9}{\includegraphics{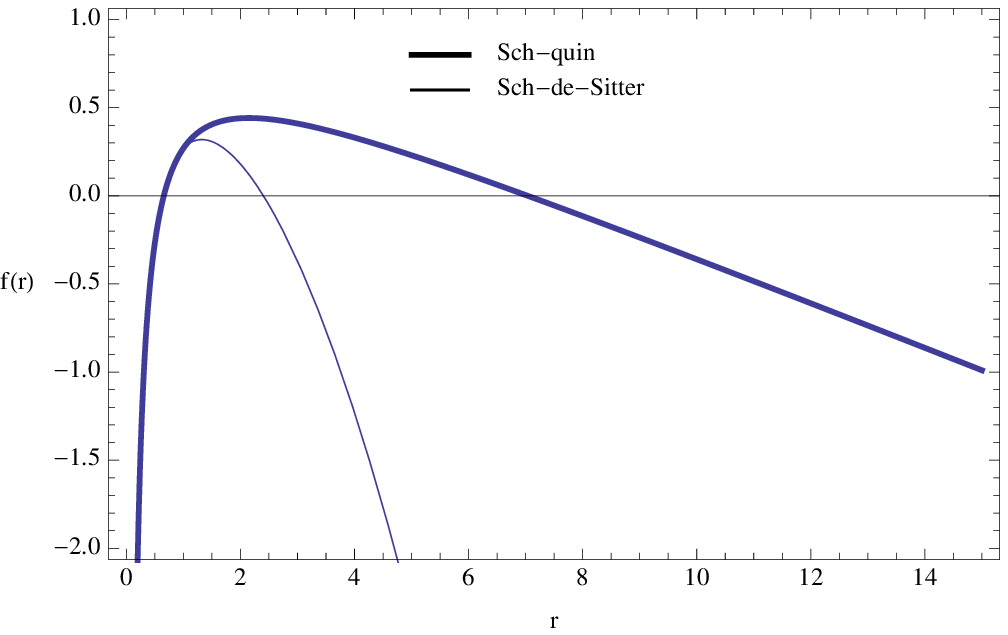}}

\vspace{0.3cm}

 \end{center}

Figure 6. The figure shows the $f(r)$ vs $r$ for Schwarzschild-quintessence black hole and the Schwarzschild-de Sitter black hole. Here, $ M = 0.3$ and $ \alpha = \frac{ \Lambda}{3} = 0.13$.\\

\noindent
One can observe that $r_{c}$ in Schwarzschild-quintessence black hole is similar to the outer horizon in Schwarzschild-de-Sitter black hole \cite{podolsky}. In order to understand how they differ, the horizons are computed and plotted with varying mass for both black holes in Fig.7 and Fig.8.

\begin{center}
\scalebox{.9}{\includegraphics{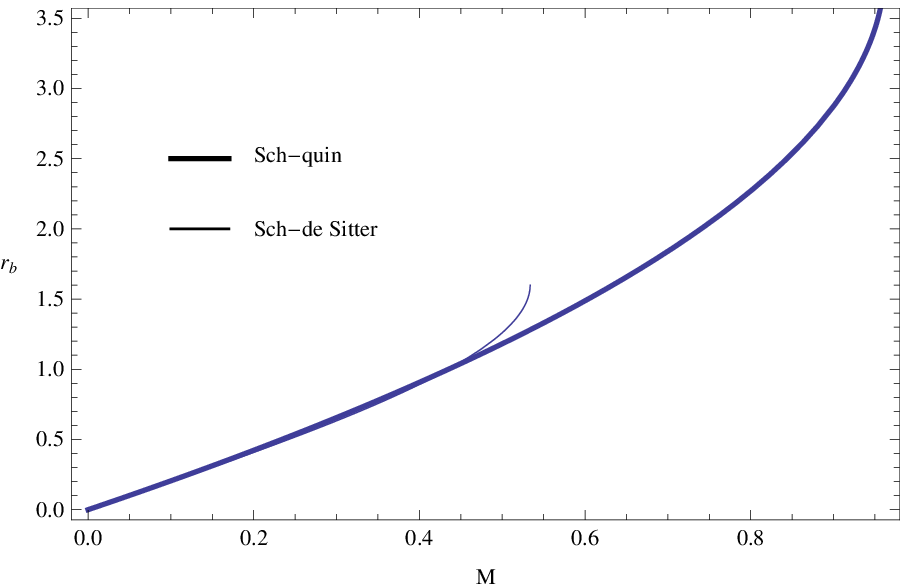}}

\vspace{0.3cm}

 \end{center}

Figure 7. The figure shows the $r_b$ vs $M$ for Schwarzschild-quintessence black hole and the Schwarzschild-de Sitter black hole. Here, $ \alpha = \frac{ \Lambda}{3} = 0.13$.\\

\begin{center}
\scalebox{.9}{\includegraphics{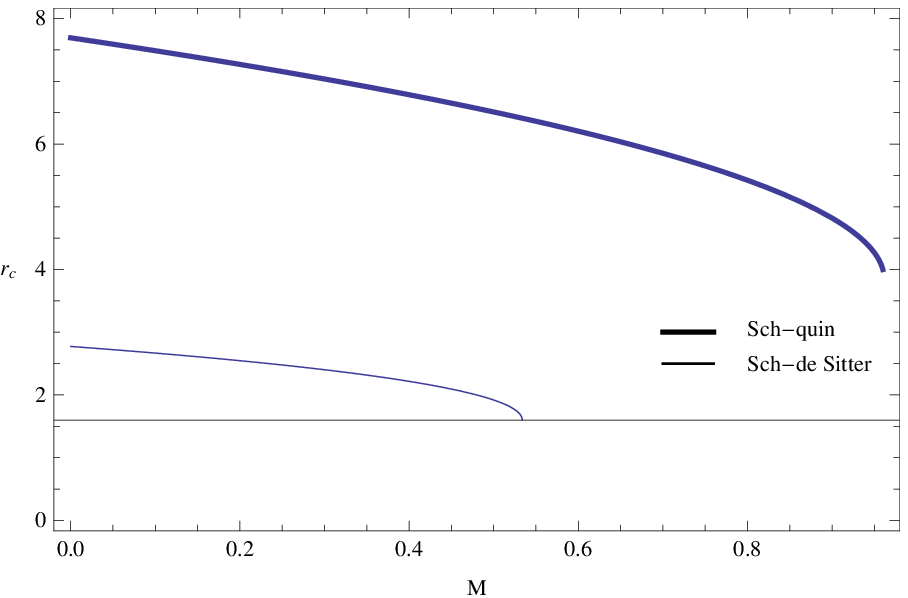}}

\vspace{0.3cm}

 \end{center}

Figure 8. The figure shows the $r_c$ vs $M$ for Schwarzschild-quintessence black hole and the Schwarzschild-de Sitter black hole. Here,  $ \alpha = \frac{ \Lambda}{3} = 0.13$.\\
 
When $ \alpha = \frac{ \Lambda}{3}$, the black hole radius is similar for both black holes. However,  Schwarzschild-de Sitter black hole  will become degenerate earlier.  The cosmological horizon is smaller  for the Schwarzschild-de  Sitter black hole.


\subsection{ Temperature}

The temperatures of the two black holes are plotted in Fig.9 and Fig.10. for comparison.

\begin{center}
\scalebox{.9}{\includegraphics{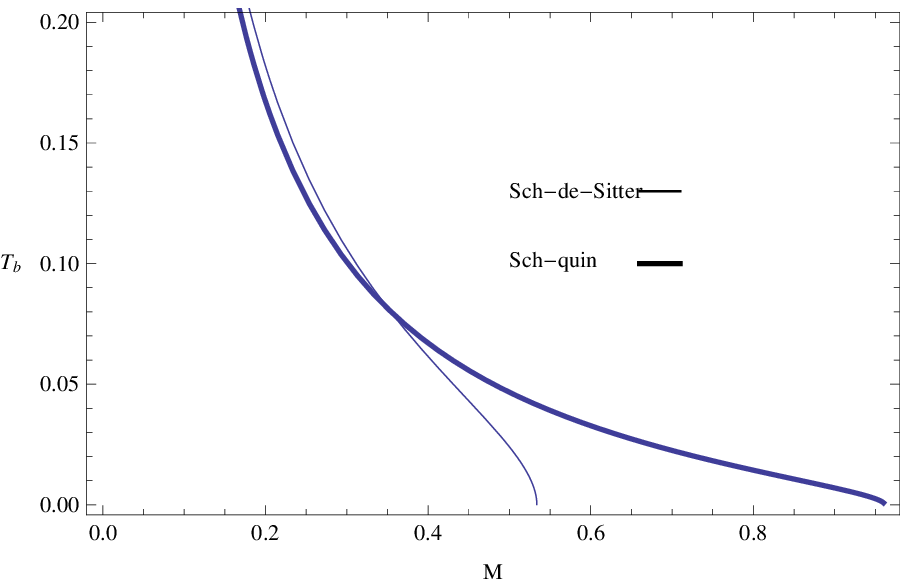}}

\vspace{0.3cm}

 \end{center}

Figure 9. The figure shows the $T_b$ vs $M$ for Schwarzschild-quintessence black hole and the Schwarzschild-de Sitter black hole. Here,  $ \alpha = \frac{ \Lambda}{3} = 0.13$.\\

\begin{center}
\scalebox{.9}{\includegraphics{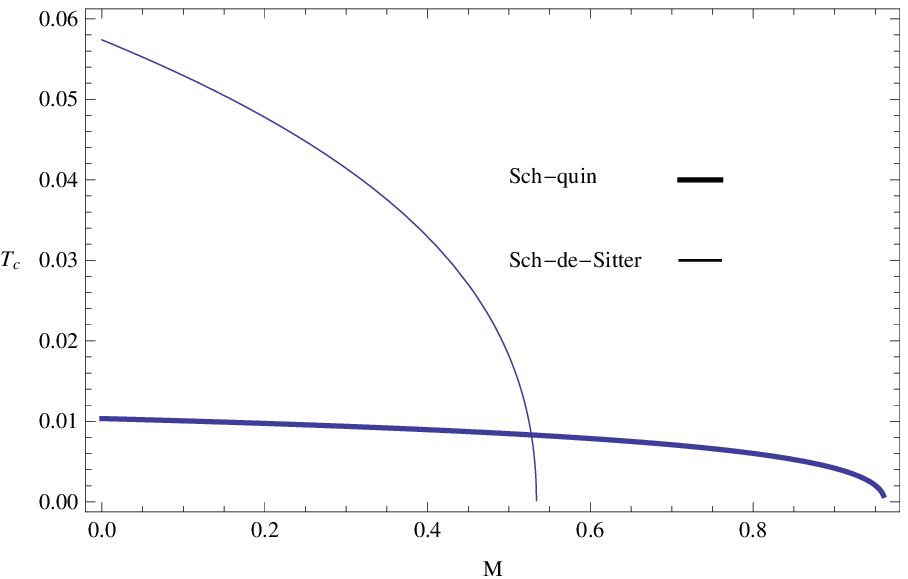}}

\vspace{0.3cm}

 \end{center}

Figure 10. The figure shows the $T_c$ vs $M$ for Schwarzschild-quintessence black hole and the Schwarzschild-de Sitter black hole. Here,  $ \alpha = \frac{ \Lambda}{3} = 0.13$.\\

\noindent
From the graph, for small mass, the Schwarzschild-de Sitter black hole is hotter. For large mass, the Schwarzschild-quintessence black hole is hotter. The cosmological horizon is hotter for the Schwarzschild-de Sitter black hole.


\subsection{ Entropy}

Since both black holes have two horizons, there are two separate entropy components defined on each horizon as,
\begin{equation}
S_b = \frac{ A_b}{4} = \pi r_b^2
\end{equation}
\be
S_c = \frac{ A_c}{4} = \pi r_c^2
\ee
How ever, the total entropy of the system is given by,
\be
S = S_b + S_c
\ee
In the Fig. 11, the total entropy for both black holes are plotted. For  small $\alpha$, which is the case supported by experiments, the Schwarzschild black hole with the quintessence has higher entropy for the same mass. Also, when $ M \ra 0$, the entropy increases in both cases. Since according to the second law of thermodynamics, all physical systems evolve in the direction of increasing entropy, both black holes will evaporate and will be stable when $ M =0$. Hence the maximum entropies for both black holes (at $ M =0$) is given by,
\be
S_{de Sittter}^{max} = \frac{ \pi \Lambda}{3}
\ee
\be
S_{quin}^{max} =  \frac{ \pi} { \alpha^2}
\ee

\begin{center}
\scalebox{.9}{\includegraphics{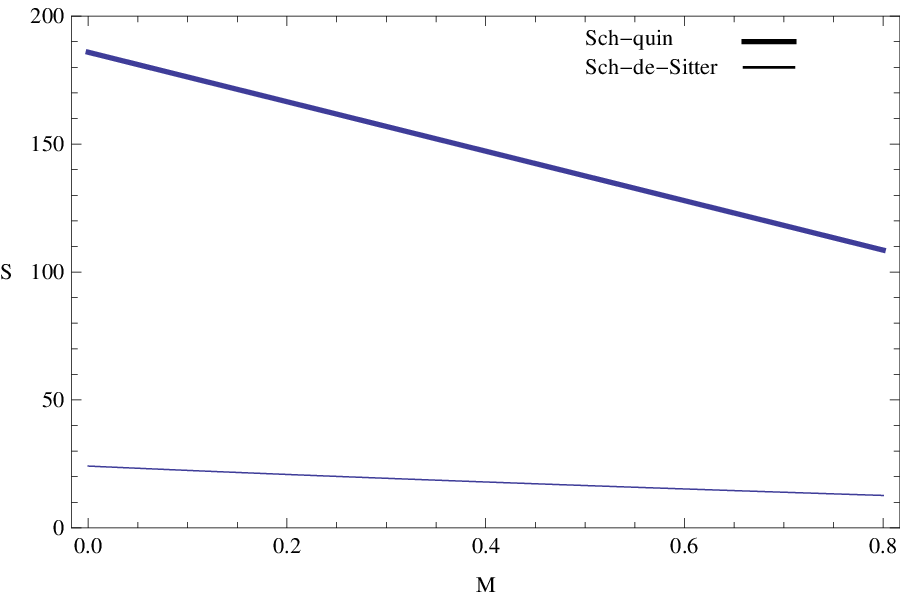}}

\vspace{0.3cm}

 \end{center}

Figure 11. The figure shows the $S$ vs $M$ for Schwarzschild-quintessence black hole and the Schwarzschild-de Sitter black hole. Here,  $ \alpha = \frac{ \Lambda}{3} = 0.13$.\\


\section{ Nariai black hole with quintessence}

In  Nariai black holes, the cosmological horizon and the black hole event horizon coincide. The extreme black hole of the Schwarzschild black hole with the quintessence has  degenerate horizons at $ r_{ex}= r_b = r_c = \frac{ 1}{ 2 \alpha}$. This   is similar to the Nariai black hole  in Schwarzschild-de Sitter space-time with the degenerate horizon $ r_{ex} =   \frac{1}{\sqrt{\Lambda}}$.  The horizon radius of the Nariai black holes are plotted in Fig.12.  The Schwarzschild-quintessence black hole has a larger radius for the Nariai case.

\begin{center}
\scalebox{.9}{\includegraphics{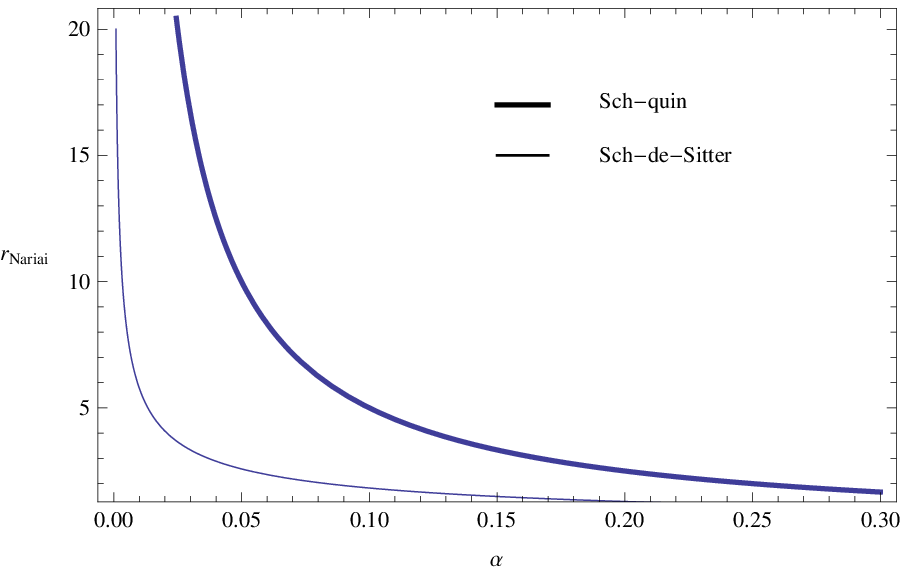}}

\vspace{0.3cm}

 \end{center}

Figure 12. The figure shows the $r_{ex}$ vs $\alpha$ for both Nariai black holes.\\

For the quintessence case, the function $f(r)_{Sch-quin}$ can be written as,
\begin{equation}
f(r)_{Sch-quin-Nariai}= - \frac{( r - r_b)^2}{ 2 r_b r}
\end{equation}

For a nearly extreme black hole of this type, one can approximate the function $f(r)$ as \cite{paw},

\begin{equation}
f(r) = \frac{ f''(r_{ex})}{2} ( r - r_1) ( r - r_2)
\end{equation}
Here, $r_1$ and $r_2$ represents the two close horizons, $r_{b}, r_{c}$. One can introduce new coordinate $ \chi$ as,
\begin{equation} \label{new}
r = r_{ex} + \delta cos \chi
\end{equation}
around the horizons which are close. Here $\delta$ is small. From eq.$\refb{new}$, $\chi=0$ corresponds to $r_b$ and $\chi = \pi$ corresponds to $r_{c}$. A new time-like coordinate $\psi$ is also introduced such as,
\begin{equation}
t = \frac{ 2 \psi}{ \delta f''(r_{ex}) }
\end{equation}
By substituting the new coordinates, the metric can be written as,
\be
ds^2 = \frac{ -2}{ f''(r_{ex})} \left( - sin^2\chi d \psi^2 + d \chi^2 \right) + r_{ex}^2 d \Omega^2
\end{equation}
The above geometry corresponds to $ dS_2 \times S^2$. The $dS_2$ has a scalar  curvature,  
\be
 R_{dS_2}=   |f''(r_{ex})|
 \ee
Since $f''(r_{ex})$ is negative due to the nature of the function $f(r)$ at $ r_{ex}$, one has to take the absolute value.

For the Schwarzschild black hole with quintessence, 
\be \label{fdouble}
f''(r_{ex}) = - 4 \alpha^2
\ee
Hence, 
\be
R_{dS_2} =  4 \alpha^2
\ee
The curvature of the $dS_2$ increases with $\alpha$. One may  recall that the Schwarzschild-de Sitter black hole also has the same topology near the degenerate horizon but with the scalar curvature, $2\Lambda$.  The scalar curvature for the Nariai black hole is   plotted in Fig.13. The curvature
increases with  quintessence parameter.

\begin{center}
\scalebox{.9}{\includegraphics{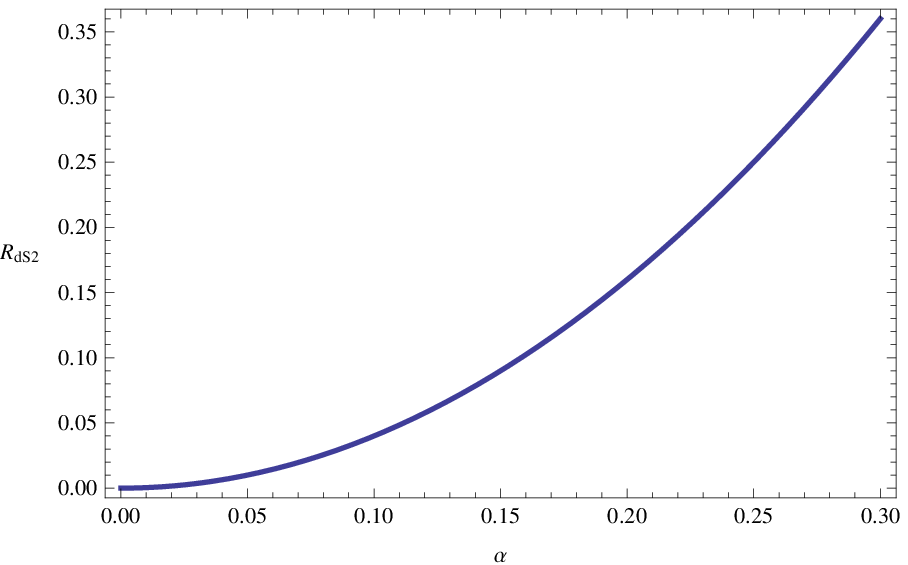}}

\vspace{0.3cm}

 \end{center}

Figure 13. The figure shows the $R_{dS2}$ vs $\alpha$ for the  Nariai black holes.\\

\subsection{ Proper distance}

Even though the horizons coincide in the Schwarzschild-like coordinates, the proper distance between them is not-zero. it can be calculated in the new coordinate system as,
\be
\int^{\pi}_{0} \frac{ \sqrt{2}d \psi}{ \sqrt{ |f''(ex)|}} = \frac{ \pi}{ \sqrt{2} \alpha}
\ee
In comparison, the proper distance between the horizons in the Schwarzschild-de Sitter Nariai case is $\frac{\pi}{ \sqrt{\Lambda}}$.

\subsection{ Temperature}

The Hawking temperature of the Nariai black hole is not zero as expected.  Cho and Nam \cite{cho} redefined the surface gravity for the Nariai-type black holes as,
\be
\tilde{\kappa}^2 = \frac{ f''(r_{ex})}{2}
\ee
Hence the temperature at the degenerate horizon is,
\be
T = \frac{\sqrt{|f''(r_{ex})|}}{ 2 \sqrt{2} \pi} = \frac{ \alpha}{ \sqrt{2} \pi}
\ee
In comparison, the temperature of the Schwarzschild-de Sitter Nariai black hole is $ \frac{ \sqrt{\Lambda}}{ 2 \pi}$. Both temperatures for the Nariai case is plotted in Fig.14. The quintessence Nariai black hole is colder than the de Sitter case.

\begin{center}
\scalebox{.9}{\includegraphics{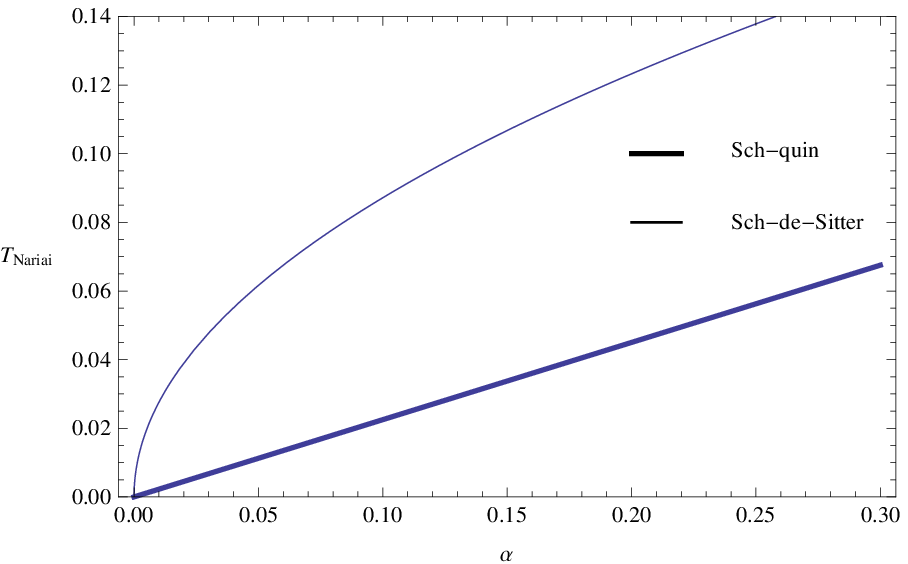}}

\vspace{0.3cm}

 \end{center}

Figure 14. The figure shows the $T$ vs $\alpha$ for both Nariai black holes.\\

\subsection{Entropy}

In the limit when the horizons coincide, the entropy of the system reaches the minimum. The entropy of the Nariai black hole for both cases are,
\be
S^{Nariai}_{quin} =  \frac{ 2 \pi}{ \Lambda}
\ee

\be
S^{Nariai}_{de Sitter} =   \frac{ \pi}{ \alpha^2}
\ee
Both entropies are plotted in Fig.15. The black hole with the quintessence has lower entropy.

\begin{center}
\scalebox{.9}{\includegraphics{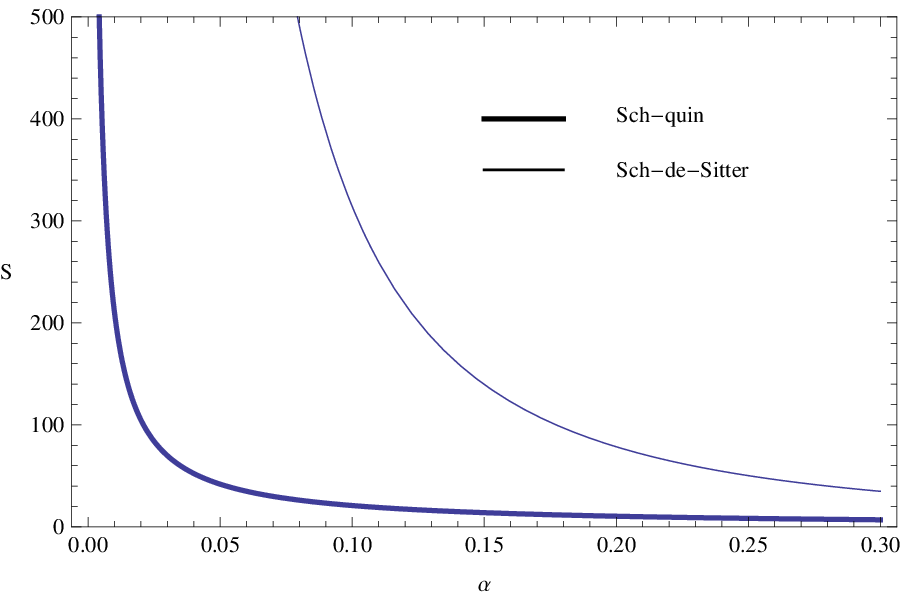}}

\vspace{0.3cm}

 \end{center}

Figure 15. The figure shows the $S$ vs $\alpha$ for both Nariai black holes.\\


\section{ Nariai black hole for general value of $\omega$ }

In this section we will briefly visit the Nariai black hole for any value of $\omega$ within the range of $ -1 < \omega < -  \frac{1}{3}$. Since both $f(r)$ and $f'(r)$ are zero for this special case, one can solve the degenerate horizon as,
\be
r_{ex} =  \frac{ 6 M \omega}{ 1 + 3 \omega}
\ee
and the value of $\alpha$ as,
\be
\alpha = - \left( \frac{ 2 M \omega}{ 1 + 3 \omega} \right)^{ 1 + 3 \omega} \frac{27 ^{\omega}}{ \omega}
\ee
For this special case,  
\be
f''(r_{ex}) = \frac{ ( 1 + 3 \omega)^3}{ 36 M^2 \omega^2}
\ee
Hence the Nariai black hole with arbitrary $\omega$ has the scalar curvature,
\be
R_{dS_2} = \frac{ | ( 1 + 3 \omega)^3|} { 36 M^2 \omega^2}
\ee
and the temperature is given by,
\be
T_{Nariai} = \frac{ 1}{ 2 \sqrt{2} \pi} \sqrt{ \frac{ | ( 1 + 3 \omega)^3|} { 36 M^2 \omega^2}}
\ee
Both the temperature and the effective curvature are plotted in Fig.16 and Fig.17. It is clear that with increasing $\omega$, both quantities decrease. Even though  these are extreme black holes, the temperature is not zero for the chosen values of $\omega$.

\begin{center}
\scalebox{.9}{\includegraphics{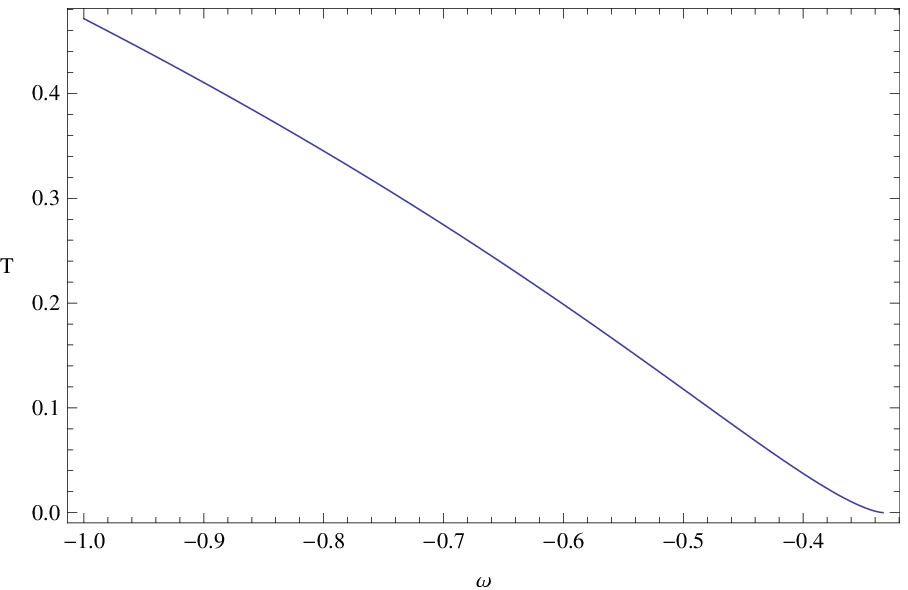}}

\vspace{0.3cm}

 \end{center}

Figure 16. The figure shows the $T$ vs $\omega$ for Schwarzschild-quintessence black hole. Here, $ M = 1$\\

\begin{center}
\scalebox{.9}{\includegraphics{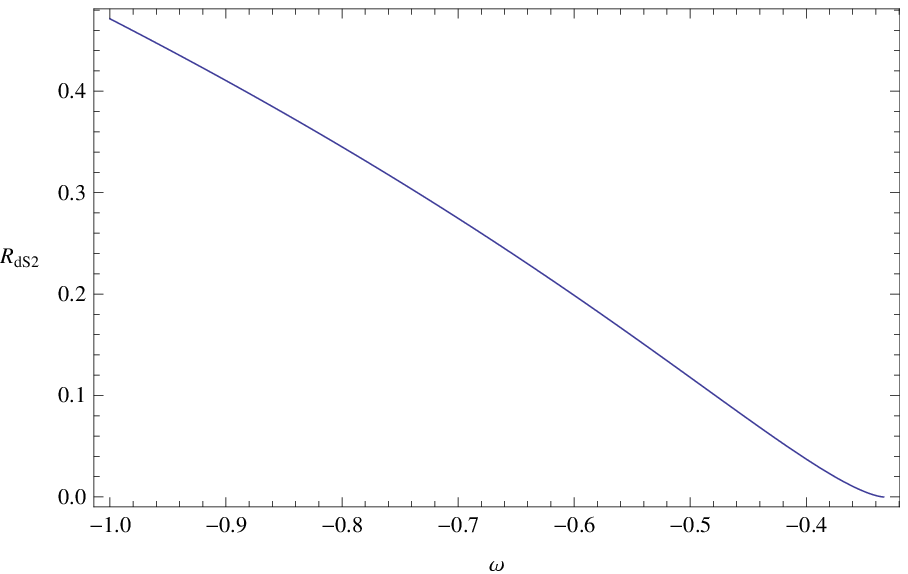}}

\vspace{0.3cm}

 \end{center}

Figure 17. The figure shows the $R_{dS2}$ vs $\alpha$ for Schwarzschild-quintessence black hole.  Here, $ M = 1$\\


\section{Conclusions}

In this paper, we have studied black hole surrounded by the quintessential matter. First we focused on the black hole with the state parameter $\omega = -\frac{2}{3}$. Such a black hole can have two horizons: one, a black hole horizon and the other a cosmological horizon. For $ M = \frac{1}{8 \alpha}$, there are degenerate black holes which are called Nariai black holes similar to the Schwarzschild-de Sitter  extreme black holes. The topology of the Nariai black hole is $dS_2 \times S_2$. The $dS_2$ has the scalar curvature $ 4 \alpha^2$. The temperature of the degenerate black hole is not zero. Its value is $\frac{\alpha} { 2 \pi}$. We have done a detailed comparison of physical properties of the black hole with the quintessence and the Schwarzschild-de Sitter black hole.

In the last section of the paper, we studied the Nariai type black hole with a general value of the state parameter $\omega$ (in the range$ -1 < \omega < -\frac{1}{3}$). Such a black hole  has a degenerate horizon at $ r = \frac{ 6 M \omega}{ 1 + 3 \omega}$. We found the temperature of the black hole and the scalar curvature of  $dS_2$.

As future work, it would be interesting to study the thermodynamics of the Nariai black hole with the quintessence along the lines of the work in  \cite{yun} \cite{eune}.

The quantum radiation of the black hole with  quintessence would be another interesting aspect to study. When $r_b$ is smaller than $r_c$, the effects of the radiation from the cosmological horizon is not significant. However, when $r_b = r_c$,  the thermal equilibrium of the solution may be unstable. A detailed study on the evaporation of the Nariai black hole with the quintessence would be an interesting aspect to investigate along the lines of the work of Bousso and Hawking \cite{raphael2}.


\end{document}